\documentclass[aps,subeqns.floatfix,twocolumn,amsmath,showpacs]{revtex4-1}
\usepackage{graphicx}
\usepackage{dsfont}
\usepackage{caption}
\usepackage{subcaption}
\usepackage{ragged2e}
\DeclareCaptionJustification{justified}{\justifying}
\begin{document}	
\title{Chimeralike states in a network of oscillators under attractive \\ and repulsive global coupling}
\author{Arindam Mishra $^{1,2}$,  Chittaranjan Hens$^{4}$, Mridul Bose $^1$, Prodyot K. Roy $^3$,  Syamal  K. Dana$^2$,}
\affiliation{$^1$Department of Physics, Jadavpur University, Kolkata 700032, India}
\affiliation{$^2$CSIR-Indian Institute of Chemical Biology, Kolkata 700032, India}
\affiliation{$^3$Department of Mathematics, Presidency University,  Kolkata 700073, India}
\affiliation{$^4$Department of Mathematics, Bar-Ilan University, Ramat Gan 529002, Israel}

 \date{\today}
 \begin{abstract}
We observe chimeralike states in an ensemble of oscillators using a type of global coupling consisting of two components: attractive and repulsive mean-field feedback. We identify existence of two types of chimeralike states in a bistable Li\'{e}nard system; in one type, both the coherent and the incoherent populations are in chaotic states (called as chaos-chaos chimeralike states) and, in another type, the incoherent population is in periodic state while the coherent population has irregular small oscillation. Interestingly, we also recorded a metastable state in a parameter regime of the Li\'enard system where the coherent and noncoherent states migrates from one to another population. To test the generality of the coupling configuration, we present  another example of bistable system, the van der Pol-Duffing system where the chimeralike states are observed, however, the coherent population is periodic or quasiperiodic and the incoherent population is of chaotic in nature. Furthermore, we apply the coupling to a network of chaotic R\"ossler system where we find the chaos-chaos chimeralike states.
\end{abstract}
\maketitle
\section*{I. Introduction}
Chimera states emerge \cite{Kuramoto, Strogatz, Martenes, Sen-2008, Sheeba-2010, Omelchenko-chaos, Omelchenko-2013, Sen-2013, Anna, Davidsen} as sequentially organized subpopulations of coherent and incoherent dynamical units in a network of oscillators under nonlocal coupling. From first observation of this unexpected phenomenon in a network of phase oscillators \cite{Kuramoto, Strogatz} in the weak coupling regime, till date it has been reported to exist in limit cycle systems \cite{Sen-2013, Omelchenko-2013} and chaotic systems \cite{Davidsen}  in the stronger coupling limit too. There in addition to phase incoherence,  amplitude incoherence of a subpopulation has been found in the chimera states. Evidence of chimers states, by this time, has been found in chemical \cite{Tinsley}, opto-electronic \cite{Murphy} and electronic circuit experiments \cite{Maistrenko, Gauthier}, and lately, in an experiment with network of metronomes \cite{Martenes-2013}. 
\par  Three different categories of chimera states have so far been identified \cite{Kapitaniak}  in networks of limit cycle or chaotic  oscillators under nonlocal coupling.  The basic chimera structure is composed of  an incoherent subpopulation in a chaotic state while the coherent subpopulation could be  periodic \cite{Kuramoto, Strogatz,  Sen-2008,  Sen-2013,  Davidsen} or remain close to a steady state \cite{Kapitaniak}. In another type of  chimera states \cite{Omelchenko-chaos, Omelchenko-2013}, the incoherent population remains in a state of spatial chaos \cite{Hassler} while the coherent population may be in a steady state or periodic state. A third kind of chimera state is classified as to coexisting structure of spatial chaos and spatio-temporal chaos \cite{Kapitaniak} in the incoherent population. At least one bistable system was found \cite{Kapitaniak} where all three types of chimera states exist in different parameter regimes, however, we emphasize that the network was under nonlocal coupling. 
  \par Chimera states are intriguing since it emerges in an  ensemble  of identical oscillators under symmetric coupling although nonlocal. It is more nontrivial in an ensemble of identical oscillators under all-to-all global coupling since no spatial sequence or identity of the oscillators exists.  However, a population of globally coupled oscillators was reported to split \cite{Sen-2014, Pikovsky, Schimdt} into synchronized and desynchronized subpopulations which has also been called as chimera states. We preferably call it chimeralike states as suggested by others \cite{Pikovsky} since there is no spatial pattern yet reminiscent of the chimera states under the nonlocal coupling. Such chimeralike states were noticed in the past \cite{Kuramoto93, Kaneko, Daido} in globally coupled network, although not defined explicitly until stated categorically by Sen et al \cite{Sen-2014}. Almost at the same time, it has also been reported in limit cycle systems for a nonlinear global coupling \cite{Schimdt}, globally coupled phase oscillators with delayed feedback \cite{Pikovsky}.  
The mechanisms of the emergence of chimeralike states differ for different coupling configurations in limit cycle systems; it is either amplitude mediated \cite{Sen-2014, Schimdt} or amplitude modulated chimera \cite{Schimdt} .  In the chimeralike states too, the phase and/or amplitude of the coherent population are randomly distributed in the incoherent population and the coherent population is in periodic state. 
\par Nonisochronicity \cite{Daido, Sen-2014, Kopell, Blasius} plays a crucial role in the chimeralike states of globally coupled network such as the case of Complex Ginzburg-Landau system \cite{Sen-2014} and the van der Pol system \cite{Hens}. Otherwise a nonlinear global coupling \cite{Schimdt} can also break the symmetry of a population into synchronous and nonsynchronous subpopulations. The presence of delay feedback  as shown in a network of phase oscillators under global coupling may also create \cite{Pikovsky} such bistability of synchronous and nonsynchronous states in a population. Alternatively, a combination of attractive and repulsive coupling was also shown \cite{Pikovsky} to break the symmetry of globally coupled bistable oscillators to create chimeralike states, however, they were forced into separate state variables of each dynamical unit of the network from an external dynamical source. 
\par In contrast, in this paper, we use the attractive and repulsive global interactions in networks of oscillators, limit cycle and chaotic systems, to observe chimeralike states but do not apply the coupling interaction from an external dynamics \cite{Pikovsky}. We assume that the coupling interactions originate within the system. We use the attractive coupling as a mean-field self-feedback while the repulsive interaction is introduced either as a mean-field self-feedback or a cross-feedback. A type of self-feedback and cross-feedback coupling was used earlier \cite{Omelchenko-2013} in a network under nonlocal coupling, where both attractive and repulsive interaction were present, to show chimera/multichimera states. We use a similar self-feedback as well as cross-feedback but purely global coupling. We first apply the coupling to a network of bistable Li\'enard system \cite{Chandrasekhar} and find two types of chimeralike states in two different regions of the parameter space of the network where the dynamics are qualitatively different in the synchronous and nonsynchronous populations. In one type, both the synchronous and asynchronous populations are chaotic, which is different from the typically observed \cite{Sen-2014, Pikovsky, Schimdt} dynamics in the chimeralike states in limit cycle systems. In the second type of chimeralike states, the noncoherent population is periodic but with no phase coherence which is qualitatively similar to the chimera states for nonlocal coupling reported earlier \cite{Omelchenko-2013}. The coherent population shows a small oscillation close to the steady state but of irregular  nature. 
 Furthermore, we report various clustered states (1-, 2-, 3-, 4-cluster) and a special kind of dynamical behavior, namely, a metastable state in the network of the Li\'enard systems. In this metastable state, both the coherent and incoherent states migrate from one subpopulation to another in time, however, we find a distinct network size effect in its transient behavior which we elaborate later.
To further exemplify the role of the proposed coupling, we apply it to another network of bistable van der Pol-Duffing system and confirm the presence of chimeralike states where the noncoherent population is typically in chaotic state while the coherent subpopulation is  periodic or quasiperiodic. Next for a network of chaotic R\"ossler systems, we simplify the coupling by separating the attractive self-feedback and the repulsive cross-feedback coupling from a single variable and apply both as self-feedback to two different variables and find clear evidence of the chimeralike states. The dynamics is typically chaotic in both the coherent and noncoherent subpopulations which we call as chaos-chaos chimeralike states.  We elaborated the coupling structure in the next section, and  we located the parameter regions of two different types of chimerlike states, the clustered states and and the metastable state in the network of Li\'enard system in section III. The chimeralike states in the van der Pol-duffing system and the  R\"ossler system are described in sections IV and V respectively. Results are summarized in  section VI.
\section*{II. Netork coupling configurtion}
The dynamics of the {\it i}-th node of the network is expressed by, ${\bf \dot{X}_i}={ F(\bf X_i,\mu)}+K AB$ where $i=1,...,N$; ${\bf\mu}$ is the set of system parameters, $K$ is the strength of coupling. All the dynamical nodes in the network are identical, $F:\bf R^2\rightarrow \bf R^2$ (considering 2D systems here,  extendable easily to higher dimension);  ${\bf X_i}=[x_i,y_i]^{T}$ where $i=1,...,N$, ${F(\bf X_i,\mu)}=[f(x_i,y_i,\mu_1,..),g(x_i,y_i,\mu'_1,...)]^T$. $A$ is a $2\times 2$ matrix with real values and $B$ is a $2\times 1$ matrix defining two types of mean-field diffusions,  
\[ A=\begin{pmatrix}
  a_{11} & a_{12} \\
   a_{21} & a_{22}
 \end{pmatrix},  
B=\begin{pmatrix}
 \bar {x}-x_i \\
 \bar {y}-y_i 
 \end{pmatrix}
\]
where $\bar {x}$=$\frac{1}{N} \sum\limits_{j=1}^N {x_j}$ and $\bar {y}$= $\frac{1}{N} \sum\limits_{j=1}^N  {y_j}$.
We discuss about different options of the proposed coupling, {\it Case }I:  $a_{11}=1$ and $a_{12}=a_{21}=a_{22}=0$ describes the conventional global coupling, a type of self-feedback acting on a scalar variable $x$. {\it Case} II:   $a_{11}= 0$,  $a_{12}= 0$, $a_{21}=\epsilon$ and $a_{22}=1$.  The global coupling now consists of two components, one self-feedback  involving the $x$-variable and another cross-feedback involving the $y$-variable; they are both added to the dynamics of the $y$-variable of each dynamical unit of the network.
Varying $\epsilon$ from $+ve$ to $-ve$ value, the cross-feedback coupling changes from attractive to repulsive nature. A combined effect of $K$ and $\epsilon$ on the collective and macroscopic behavior of the whole network is investigated.
{\it Case} III:  $a_{11}=1$ and $a_{22}=\epsilon$, other two elements are zero. The global coupling is established by applying two mean-field self-feedback interactions to the dynamics of two different state variables of each unit. {\it Case} IV: all the elements in matrix $A$ are nonzero when it is of complex type. We focus here on the {\it Cases} II-III and explore chimeralike states in different example systems. Note that we adopted a similar global coupling earlier \cite{Hens} to observe the chimeralike states in the van der Pol system and the chaotic R\"ossler oscillator. We simplify the coupling here and show emergence of the chimeralike states in the bistable Li\'enard system, and a bistable van der Pol-Duffing system and the R\"ossler system. Especially, we make elaborate studies on the network of the Li\'enard system and, locate  regions of clustering, two different kinds of chimeralike states and a type of metastable states in parameter space.
\begin{figure*} []
\centering
\includegraphics[width=\textwidth]{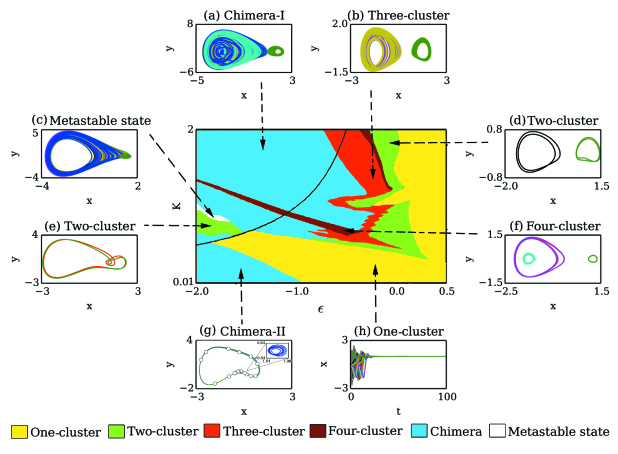}
\caption{Phase diagram in $K-\epsilon$ plane. Different dynamical states are denoted as 1-cluster (dark red),  2-cluster (pink), 3-cluster (red), 4-cluster (blue), chimeralike states (yellow). Various dynamics are shown at different parameter regimes: attractors of (a) Chimera-I ($K=1.6, \epsilon=-1.8$), (b) 3-cluster ($K=1.7, \epsilon=-0.47$), (c)  metastable state ($K=0.91, \epsilon=-1.95$), (d) 2-cluster ($K=1.7, \epsilon=-0.09$), (e) 2-cluster ($K=0.7, \epsilon=-1.8$), (f) 4-cluster ($K=0.7, \epsilon=-0.59$), (g) Chimera-II ($K=0.3, \epsilon=-1.8$), time series of all the oscillators in (h) 1-cluster state ($K=0.2, \epsilon=-0.2$).} 
\label{FIG.1}
\end{figure*}
\section*{III. Chimeralike states: Li\'{e}nard system}
\par We start our numerical study with an example of a Li\'{e}nard system \cite{Chandrasekhar} and form a network of $N$ number of identical units using the {\it Case} II  coupling, 
\[ A=\begin{pmatrix}
  0 & 0 \\
   \epsilon & 1
 \end{pmatrix};
  B=\begin{pmatrix}
    \bar{x}-x_i\\
    \bar{y}-y_i \\
    \end{pmatrix}\]. 
The globally coupled network of the Li\'enard system is
\begin{eqnarray}
\dot{x_i}&=&y_i ,\\ 
\dot{y_i}&=&-\alpha x_i y_i-\beta x_i^{3}-\gamma x_i +K[(\bar {y}-y_i) +\epsilon (\bar {x}-x_i)]. 
\end{eqnarray}
The Li\'enard system shows bistablity in isolation \cite{LeoKingston}: for a choice of parameters, a stable focus coexists with periodic orbits. More categorically, for a choice of system parameters as given below, the system has a saddle separatrix at  $(0,0)$ between a stable focus at  $(1,0)$ and a saddle focus at $(-1, 0)$. A homoclinic orbit (HO) exists at the saddle point $(0,0)$, which separates the state-space into two regions. For choice of initial conditions inside the HO, the system moves to the  stable focus $(1,0)$ after a transient; for choice of initial conditions outside the HO, multiple periodic orbits appear  which have different frequencies, i.e, the system behaves like a non-isochronous system \cite{Kopell, Blasius, Daido}.
\par Next we draw a phase diagram to demarcate the parameter regions of different macroscopic dynamics of the network, coherent or noncoherent states, two different chimeralike states using a statistical measure \cite{Gopal}, namely, a strength of incoherence ($S$). 
For this measure, the whole population is divided into $M$ number of bins of equal length $n = N/M$ and a local standard deviation $\sigma(m)$ is then defined 
\begin{equation}
\sigma(m) = \left\langle\sqrt{\frac{1}{n}\sum_{j=n(m-1)+1}^{nm}[z_{j}-\langle z\rangle]^{2}}\right\rangle_{t}
\end{equation}
where $m = 1, 2,..., M$, $\langle z\rangle = \frac{1}{N}\sum_{i=1}^{N}z_{i}(t)$, $z_{i} = x_{i} - x_{i+1}$ and $< . >_t$ defines a time average. Using this local standard deviation, we measure the strength of incoherence ($S$) as 
\begin{equation}
S = 1 - \dfrac{\sum_{m=1}^{M}s_{m}}{M}, 
s_{m} = \Theta(\delta - \sigma (m))
\end{equation}
where $\Theta (.)$ is the Heaviside step function, and $\delta$ is a small predefined threshold.
Chimera states and coherent states are distinguished by the $S$ value, in general, for nonlocal and global coupling. The $S = 0$ identifies the incoherent state while $S = 1$ defines the coherent state. The $S$ lies in-between  $0<S<1$ for chimera states.  For global coupling as mentioned above, no spatial identity or index exists for the dynamical nodes and hence the multichimera-like states, although appears, cannot be distinguished from the chimera states. What really matters in the chimeralike states in globally coupled network is the symmetry breaking of a population into  coherent and  non-coherent subpopulations. Once the coherent state is identified using the above statistical measure, we use another algorithm to separate different clustered states: we record the instantaneous value of the $x_i$ variable of all the oscillators after a long transient, and consider any two of them as belonging to a particular group wherever they are identical  to each other within a small bound. Thus the oscillators having identical $x_i$ values forms a group; each group forms a separate cluster. Finally all such separate  groups  determine the number of clusters (1-,2-,3-,4- cluster). In a clustered state, the sum of the number of dynamical units in all the groups is the total number of units in the network. If this condition fails we do not consider formation of the cluster.  We further check the temporal evolution of all the oscillators for visual check. The metastable state is first identified, by the statistical measure, as the only noncoherent state  in the parameter space. The dynamics in this regime is then looked into to recognize its unknown behavior as described below. 
\par Figure 1 shows the phase digram in $K-\epsilon$ parameter plane showing different dynamical regimes. We change $K$ and $\epsilon$ both in steps of 0.01 and use the fourth order Runge-Kutta algorithm to integrate the system with a time step size 0.01. The initial states for $y_{i}$ are chosen as $y_{i0} = 2(1 - \frac{4i}{N})$ for $1 \leq i \leq \frac{N}{2}$ and  $y_{i0} = 2(\frac{4i}{N} - 3)$ for $\frac{N}{2} + 1 \leq i \leq N$  with added small random fluctuations. All initial states for $x$-variables are set at zero. A black curve divides the $K-\epsilon$ phase space into two regions.  The network has trivial equilibrium points ($x^*_i$,$y^*_i$) at (-1,0), (0,0) and (1,0). In the region below this black curve,  the homogeneous steady state at $(1,0)$ is stable while $(-1,0)$ and $(0,0)$ are always unstable. Different clustered states coexist in this region as noted in the diagram. The emergent homogeneous steady state is consistent with earlier results \cite{Hens1} in a network under mixed attractive and repulsive global coupling. On the upper  side of the black curve, the  $(1,0)$ fixed point  becomes unstable besides the unstable fixed points $(-1,0)$ and $(0,0)$ and there, all the states, clustered or chimera states, are robust to the choice of initial conditions. Different dynamical states are shown with their phase portraits in different regions of the parameter space : 1-cluster in yellow, 2-cluster in green, 3-cluster in red, 4-cluster states in dark red. 
In two different regions of 2-cluster states, the nature of the dynamics are different although periodic. The chimeralike states are observed in the parameter space indicated by the blue regions. Above the black curve, the chimeralike states (Chimera-I in Fig. 1) are chaotic both in noncoherent and coherent populations which feature is uncommon in limit cycle systems; most importantly, it is independent of the choice of initial conditions. Similarly, above the black curve, we find strips of 2-cluster (green), 3-cluster (red)and 4-cluster states (dark red) independent of initial conditions. We find a noncoherent region (white) there which we identify later as metastable state. On the other hand, the chimeralike states (Fig. 1g) in the region below the black curve are limited to specific choices of initial conditions. There the dynamics of the noncoherent population is periodic when all the oscillators are distributed in along the trajectory (open circles in the phase portrait). As a result the noncoherent population has no amplitude and phase correlation while the coherent population is limited to small oscillations close to the steady state $(1,0)$ but of irregular nature. Different clustered states (1-, 2-, 3-, and 4-cluster) are present below the black curve and coexist with the stable focus at (1,0). Note that  no 1-cluster state exists above the black curve. 
\par To draw the stability line (black curve) that delineates the parameter space  into two regions, we analytically calculate the determinants and trace of the Jacobian matrix($J$) at the trivial equilibrium points $(-1,0)$, $(0,0)$ and $(1,0)$ in the $K-\epsilon$ parameter plane. We find that the unstable foci $(-1,0)$ and $(0,0)$ remain unstable for the whole $K-\epsilon$ phase space, only the stable focus $(1,0)$ changes its stability when crossing the black line curve. The focus (1,0) becomes unstable when crosses the stability line to the upper region, the determinant  $J$($det(J)$) is
\begin{eqnarray*}
det(J) = \lambda_{1}.\lambda_{2}...\lambda_{N} > 0
\end{eqnarray*}
and we calculate
\begin{eqnarray*}
	det(J) = (-1)^{N}[a + (N-1)\frac{K}{N} \epsilon](a -\frac{K}{N} \epsilon)^{N-1} > 0 
\end{eqnarray*}
so that, $(-1)^{N}[a + (N-1)\frac{K}{N} \epsilon](a -\frac{K}{N} \epsilon)^{N-1} > 0$.\\
\\But $a + (N-1)\frac{K}{N} \epsilon = -1$ and it implies 
\begin{eqnarray}
(-1)^{N}(a -\frac{K}{N} \epsilon)^{N-1} < 0\\
K\epsilon > -1
\end{eqnarray}
Hence the black curve in the $K-\epsilon$ phase space is a rectangular hyperbola satisfying the equation $K\epsilon = -1$ (see APPENDIX for details). For $K\epsilon > -1$, $(1,0)$ is a stable fixed point below the black curve and for $K\epsilon < -1$ it is unstable above. The black curve is numerically verified by calculating the eigenvalues of the Jacobian of the network at $(1,0)$. It is interesting to note here that the black curve is independent of the network size and it indicates that the chimeralike states are independent of the network size. The $K\epsilon < -1$ condition implies $K$ and $\epsilon$ are to be of opposite sign, i.e., a combination of attractive and repulsive coupling is necessarily to be chosen for the chimeralike states to emerge. 
\par Figure 2 presents a snapshot in polar coordinates and the spatio-temporal evolution of the $x$-variables for  chimera-I state when $K = 1.6$ and $\epsilon = -1.8$.  The synchronized population is clubbed into a square (red) shown in Fig. 2(a) while the  circles (blue) represent the desynchronized population distributed in both amplitude and phase. Both the synchronized and desynchronized populations are in chaotic state but the attractors are separated in phase space as shown in their phase portrait in Fig. 1(a). Figure 2(b) represents the time evolution of the $x_i$-variable of all the oscillators showing strips of coherent and noncoherent nodes which continue for a long run along the $y$-axis. Multiple strips of noncoherent nodes are seen which should not be confused with multichimera states as explained above. Basically the whole population breaks into two coherent and noncoherent subpopulations with no positional identity. \\

\begin{figure}[h!]
	\centering
	\hspace{-70pt}
	\begin{subfigure}[b]{0.5\columnwidth}
		\centering
		\includegraphics[width = 4cm,height = 4cm]{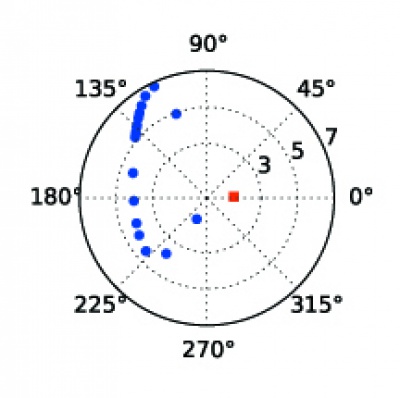}
		\subcaptionbox{\label{FIG.2}}[0.5cm]{}
		\label{fig7}	
	\end{subfigure}
	\hspace{-10pt}
	\begin{subfigure}[b]{0.5\columnwidth}
		\centering
		\includegraphics[width = 6cm,height = 3.5cm]{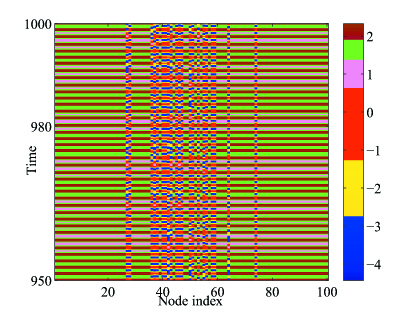}
		\subcaptionbox{\label{FIG.2}}{}
		\label{FIG.2}	
	\end{subfigure}
	\caption{Chimera-I state in a network of Li\'enard system for $K = 1.6$ and $\epsilon = -1.8$. (a) Snapshot in polar coordinate of all the oscillators, synchronized population in square (red) and nonsynchronous population in circles (blue), (b) temporal dynamics of all the nodes in the network. }
	\label{fig3}
\end{figure}

\begin{figure}[h!]
\centering
\hspace{-50pt}
\begin{subfigure}[b]{0.5\columnwidth}
\centering
\includegraphics[width = 4cm,height = 4cm]{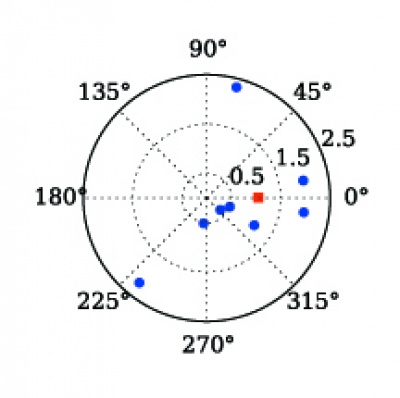}
\subcaptionbox{\label{FIG.3}}[0.5cm]{}
\label{}	
\end{subfigure}
\hspace{-10pt}
\begin{subfigure}[b]{0.5\columnwidth}
\centering
\includegraphics[width = 6cm,height = 3.5cm]{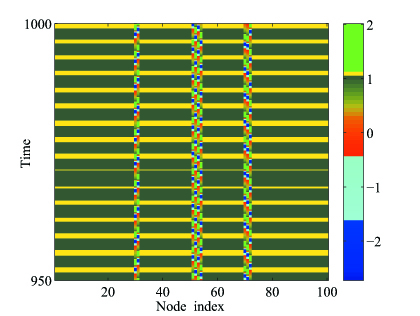}
\subcaptionbox{\label{FIG.3}}{}
\label{}	
\end{subfigure}
\caption{Chimera-II state in the network of Li\'enard system for $K = 0.3$ and $\epsilon = -1.8$. (a) snapshot in polar coordinate of all the oscillators, synchronized population in square (red) and nonsynchronous population in circle (blue), (b) temporal dynamics of all the nodes in the network. }
\label{FIG.3}
\end{figure}

\par Figure 3 shows the chimera-II state for $K = 0.3$ and $\epsilon = -1.8$. Figure 3(a) shows a snapshot in polar coordinates where the synchronized population is again denoted by a single square (red) and the desynchronized oscillators in circles (blue) as seen distributed in phase and amplitude. Figure 3(b)  shows the temporal dynamics of all the nodes which clearly reveals the coherent and noncoherent population for a long time run. The attractors of the synchronized and desynchronized regions are shown in Fig. 1(f). The desynchronized oscillators are all periodic, however, they do not have any phase coherence as seen in distributed circles denoting positions of the oscillators in 2D phase portrait. The dynamics of the synchronized population in a small dot inside the phase portrait is enlarged in the inset that shows small irregular oscillation although remains very close to the steady state at (1,0).
\par Most interestingly, an unknown kind of dynamics which we denote as metastable states appear near the boundary  of the noncoherent state (white region) and the chimera-I state in $K-\epsilon$ phase space (Fig. 1). Both the coherent and noncoherent populations are chaotic and their attractors are overlapping each other in state space (Fig. 1c). In the perspective of cortical dynamics \cite{Kelso}, a relative coordination in  neuron population has a temporal behavior;  stronger coordination at one time and becomes weaker at another time and this relative coordination may switch within the population.  Figure 4 shows the time evolution of $x$-variable of all the oscillators in the metastable states for  $K = 0.91$ and $\epsilon = -1.95$ for N=10 (left panels) and N=100 (right panels) respectively. The coherent population migrates randomly between the population in time as shown in the spatio-temporal plot in the left lower panel, which is defined as a metastable state. Coherence in the sense of coordination between a group of oscillators shows a temporal change. However, after a long transient, it moves to a cluster state as shown in the left upper panel when the number of oscillators is considered as N=10.  For N=100, the metastable states clearly continues for a long time as seen in the right panels. We find that this state is a long lived transient and the transient time increases rapidly with the oscillator number. 
We plan to explore this metastable states, in further details, in the future.\\ 

\begin{figure}[h!]
	\centering
	\hspace{-10pt}
	\begin{subfigure}[b]{0.5\columnwidth}
		\centering
		\includegraphics[width = 5cm,height = 5cm]{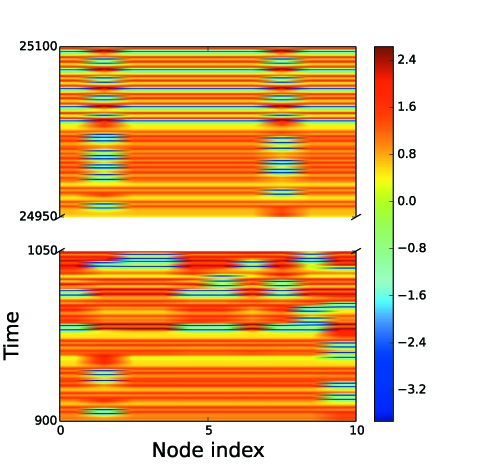}
		\subcaptionbox{\label{FIG.4}}{}
		\label{}	
	\end{subfigure}
	\hspace{-1pt}
	\begin{subfigure}[b]{0.5\columnwidth}
		\centering
		\includegraphics[width = 5cm,height = 5cm]{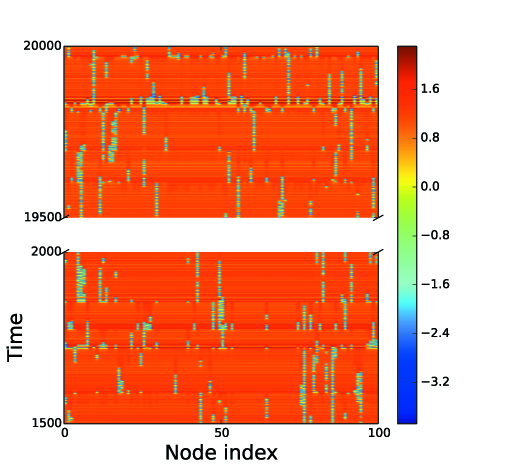}
		\subcaptionbox{\label{FIG.4}}{}
		\label{}	
	\end{subfigure}
	\caption{Temporal dynamics of $x_{i}$ variable for all the oscillators for $K = 0.91,\epsilon = -1.95$. Clustering is seen in the left upper panel ($N=10$) after a long time. Right panels ($N=100$) show no clustering. Upper and lower panels are splitted in time to show 
    the dynamics for long time in both the examples.}
	\label{FIG.4}
\end{figure}

\section*{IV. Chimeralike states: van-der Pol-Duffing system. }
We construct a network of a bistable van der Pol-Duffing system \cite{Kapitaniak} using the attractive self-feedback and repulsive cross-feedback global coupling ($Case$ II).  
\begin{eqnarray}
\dot{x_i} &=& y_i ,\\ 
\dot{y_i} &=& \alpha (1-x_i^{2}) y_i-x_i^{3}+F sin \omega t \nonumber\\
&& +\> K[\epsilon (\bar {x}-x_i)+ (\bar {y}-y_i)]. 
\end{eqnarray}
and parameter values are $\alpha = 0.2$, $F = 1$ and $\omega = 0.94$ when the isolated system is bistable having one periodic and one chaotic attractor.
The initial states for $y_{i}$ are chosen as $y_{i0} = 3(1 - \frac{4i}{N})$ for $i = 1$ to $\frac{N}{2}$ and  $y_{i0} = 3(\frac{4i}{N} - 3)$ for $i = \frac{N}{2} + 1$ to $N$ and the initial states for $x_{i}$ are chosen as $x_{i0} = 2(1 - \frac{4i}{N})$ for $i = 1$ to $\frac{N}{2}$ and  $x_{i0} = 2(\frac{4i}{N} - 3)$ for $i = \frac{N}{2} + 1$ to $N$ with an added small random fluctuation. Figure 5 shows a snapshot in polar coordinates (upper left) where the  square (red) represents the synchronized population and the distributed circles (blue) correspond to the desynchronized oscillators. In the chimeralike states, the synchronized population is in periodic state (could be quasiperiodic for a different choice of parameters) and the desynchronized oscillators are in chaotic states. Right upper panel  shows the time evolution of $x$-variables of all the oscillators which confirms the coexistence of coherent and noncoherent subpopulations and the number of oscillators in each population remain unchanged for a long run. Again we ignore the concept of multichimera for reasons explained above although the typical signature of multichimera as found for nonlocal coupling exists here. Lower panels again show two clustered state for $K=0.4$ and $\epsilon=-4$ when the left panel clearly identifies two clustered populations in a circle (blue) and a square (red) in polar coordinate and time evolution plots of all the oscillators. A small change in the $K$-value originates the chimeralike states.\\
\begin{figure}[h!]
	\centering
	\hspace{-40pt}
	\begin{subfigure}[b]{0.5\columnwidth}
		\centering
		\includegraphics[width = 4cm,height = 4cm]{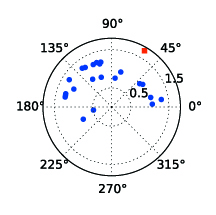}
		\subcaptionbox{\label{FIG.5}}[0.5cm]{}
		\label{}	
	\end{subfigure}
	\hspace{-10pt}
	\begin{subfigure}[b]{0.5\columnwidth}
		\centering
		\includegraphics[width = 6cm,height = 3.5cm]{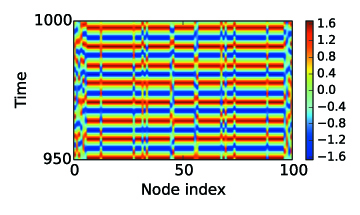}
		\subcaptionbox{\label{FIG.5}}{}
		\label{}	
	\end{subfigure} \\
	\hspace{-40pt}
	\begin{subfigure}[b]{0.5\columnwidth}
		\centering
		\includegraphics[width = 4cm,height = 4cm]{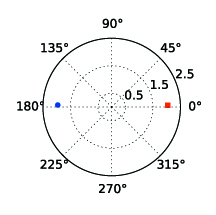}
		\subcaptionbox{\label{FIG.5}}[0.5cm]{}
		\label{}	
	\end{subfigure}
	\hspace{-10pt}
	\begin{subfigure}[b]{0.5\columnwidth}
		\centering
		\includegraphics[width = 6cm,height = 3.5cm]{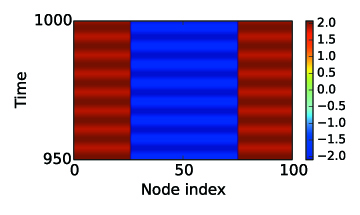}
		\subcaptionbox{\label{FIG.5}}{}
		\label{}	
	\end{subfigure}
	\caption{Chimera state for $K = 0.2$ and $\epsilon = -4$ and two cluster state for $K = 0.4$ and $\epsilon = -4$ in van der-Pol-Duffing system. Polar plots at left, upper panel shows random distribution of phase and amplitude of the nonherent oscillators in circles (blue) and single coherent cluster in square (red). Right panels plot spatio-temporal dynamics. Upper panel shows chimeralike state when all the nodes splits into coexisting coherent and noncoherent subpopulations. Lower right panel shows two  clustered populations. }
	\label{FIG.5}
\end{figure}

\begin{figure}[h!]
	\centering
	\hspace{-25pt}
	\begin{subfigure}[b]{0.5\columnwidth}
		\centering
		\includegraphics[width = 5cm,height = 3.5cm]{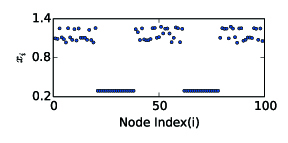}
	    \subcaptionbox{\label{FIG.6}}{}
		\label{}	
	\end{subfigure}
	\hspace{5pt}
	\begin{subfigure}[b]{0.5\columnwidth}
		\centering
		\includegraphics[width = 5cm,height = 3.5cm]{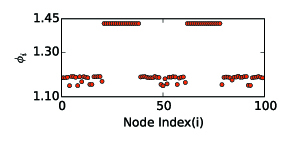}
		\subcaptionbox{\label{FIG.6}}{}
		\label{}	
	\end{subfigure}
	\caption{Chimera state for R\"{o}ssler oscillators for $K_{1} = 0.071$ and $K_{2} = 0.151$.  Snapshot of $x_i(t)$ at left and phase $\theta_i$ at right for all the oscillators. }
	\label{FIG.6}
\end{figure}
It is worth mentioning that both the attractive and the repulsive mean-field coupling may be applied as self-feedback, in the sense, that they are added separately to their corresponding state variables; the cross-global feedback is not a necessary condition to observe chimeralike states in the example systems. We do not present the results here for the systems, discussed above, however, elaborate this using the chaotic R\"ossler model. 
 
\section*{V. Chimeralike states: R\"{o}ssler system}
We simplify the coupling \cite{Hens} by separating the attractive and the repulsive mean-field interactions and apply them as self-feedback to two different dynamical equations ({\it Case} III). We choose a network of  R\"{o}ssler oscillators 
\begin{eqnarray}
\dot{x_i}&=& -y_i - z_i - K_1(\bar {x}-x_i)\\ 
\dot{y_i}&=&x_i + ay_i + K_2(\bar {y}-y_i)\\ 
\dot{z_i}&=&bx_i + z_i(x_i - c)
\end{eqnarray}
and the parameter values as $a = 0.36$, $b = 0.4$ and $c = 4.5$ in the chaotic regime and here the isolated system is not bistable.
Figure 6 shows snapshots of $x_i$-variable and the phase $\phi_i$ for $N=100$ oscillators. Left panel is a snapshot of amplitude and the right shows snapshot of phase of all the oscillators (nodes) showing signature of 
chimeralike states. The distribution of phase and amplitude along the nodes confirms the state of incoherence in one subpopulation  while the other counterpart remain synchronized. We emphasize once again that it should not be confused with multichimera states. The whole population splits into coherent and noncoherent subpopulations with no specific spatial structure.\\

\section*{VI. Conclusion} 
We observed chimeralike states in networks of identical nonlinear oscillators using a global coupling consisting of both attractive and repulsive mean-field feedback. Historically,  the chimera states were observed \cite{Kuramoto}, as a strange phenomenon, in network of identical oscillators under range limited interaction, the so called nonlocal coupling. The chimera states appeared highly nontrivial in globally coupled identical oscillators, however, the chimeralike states were reported in globally coupled network of oscillators. A  homogeneous network of globally coupled oscillators splits into coexisting coherent and noncoherent subpopulations in a selected parameter space and for strong coupling. The coupling scheme could be established as a simple global mean-field interaction but the presence of nonisochronicity was found crucial for chimera states to observe \cite{Sen-2014, Hens} in the limit cycle systems such as the Complex Ginzburg-Landau system, the van der Pol system.  Alternatively, a nonlinear coupling \cite{Schimdt} or a delay feedback \cite{Pikovsky} was used for chimeralike states to emerge in globally coupled network. A bistabilty in the dynamical units was found to augment the emergence of chimera states in limit cycle systems. However, for chaotic systems, neither the non-isochronicity nor bistability criterion is a necessary for the origin of chimeralike states as first shown in globally coupled chaotic map \cite{Kaneko}. We showed here that a combination of attractive and repulsive mean-field coupling can produce chimeralike states in a broad parameter space of a Li\'enard  system. Interestingly, we identify two different types of chimeralike states. In one type, we found both the subpopulations in chaotic state. In another type of chimeralike states, the coherent population remains close to a steady state but with a small irregular oscillation. On the other hand, the noncoherent population remains in periodic or quasiperiodic state. Thus, in the globally coupled network under two competitive coupling components, a rich variety of chimeralike states with diverse dynamical features  were found. We established the role of two competing coupling components in creating chimeralike states using numerical examples of two bistable limit cycle systems, namely, a Li\'enard system and the van der Pol-Duffing system, and the chaotic R\"ossler system. We noted that such attractive and repulsive coupling were used \cite{Pikovsky} in globally coupled chaotic systems recently to observe chimeralike states but they were forced into the network from an additional external dynamics. In contrast, we explored the chimeralike states in both limit cycle and chaotic systems where the coupling interactions are all internally generated  and applied either as a self-feedback and/or cross feedback. To make a clear distinction of the chimera states in globally coupled network from the nonlocally coupled network, we prefer the terminology such as the chimeralike states throughout the text. This is due to the fact that chimeralike states refers to a phenomenon of symmetry breaking of a homogeneous population into subpopulations of coherent and noncoherent oscillators; no spatial identity exists. On the other hand, in the traditional chimera states, a clear spatial identity of oscillators exists due to the nonlocal nature of the coupling. The whole population splits into coherent and noncoherent oscillators but organized in a spatial order. 
\par Besides chimeralike states, we observed different clustered states (1-, 2-, 3-, 4-cluster) and, most importantly, a kind of metastable state in a parameter region near the transition from a cluster state to the chimeralike states via a noncoherent state. In the metastable states, the coherent/noncoherent population migrates in time to different population of oscillators. At one instant, some oscillators were in coherent state which loses coherence with time but another population establishes coherence by that time. This migrating coherent or noncoherent state led to a permanent coherent state after a long transient for  network of smaller size ($N=10$). But this transient behavior showed a network size effect. For a reasonably larger network size ($N=100$), we could not predict the transient time of the coherence state even for large simulations within our limited computational facility. We plan to explore the metastable state, in further detail, in the future.
\par S.K.D. and P.K.R. acknowledge individual supports by the CSIR Emeritus Scientist Scheme, India. A. Mishra is supported by the University Grant Commission, India.\\

\section*{APPENDIX}
For coupled Li\'enard system the Jacobian matrix at$(1,0)$ can be written as
\begin{equation}
	J = \mathds{1}\otimes U + \mathds{I}\otimes(U-V)
\end{equation}
where \[ U=\begin{pmatrix}
0 & 1 \\
a & b
\end{pmatrix};
V=\begin{pmatrix}
0 & 0\\
\frac{K}{N}\epsilon & \frac{K}{N} \\
\end{pmatrix}\] 
with $a = -3\beta -\gamma + \frac{K}{N}\epsilon - K\epsilon$ and $b = -\alpha + \frac{K}{N} - K$.\\$\mathds{1}$ is the unit matrix of order $N\times N$ and $\mathds{I}$ is the identity matrix of order $N\times N$.\\
So the Jacobian matrix($J$) at $(1,0)$ becomes a $2N\times 2N$ square matrix and it is denoted as \\
\[ J=\begin{bmatrix}
0      & 1     &  0     & 0      & 0     &  0  & \cdots  & \cdots & 0  & 0\\\\
a      & b     &  \frac{K}{N} \epsilon   & \frac{K}{N}      & \frac{K}{N} \epsilon     & \frac{K}{N}     & \cdots  & \cdots & \frac{K}{N} \epsilon & \frac{K}{N} \\\\
0      & 0     &  0     & 1      & 0     &  0  & \cdots  & \cdots & 0 & 0\\\\
\frac{K}{N} \epsilon    &\frac{K}{N}     & a   & b     & \frac{K}{N} \epsilon     & \frac{K}{N}     & \cdots  & \cdots & \frac{K}{N} \epsilon & \frac{K}{N} \\\\
0      & 0     &  0     & 0      & 0     &  1  & \cdots  & \cdots & 0 & 0\\\\
\vdots & \vdots & \vdots & \vdots & \vdots & \vdots & \ddots & \ddots &  \vdots &  \vdots \\\\
\vdots & \vdots & \vdots & \vdots & \vdots & \vdots & \ddots & \ddots &  \vdots &  \vdots \\\\
\frac{K}{N} \epsilon     & \frac{K}{N}     &  \frac{K}{N} \epsilon   & \frac{K}{N}      & \frac{K}{N} \epsilon     & \frac{K}{N}     & \cdots  & \cdots & a & b
\end{bmatrix}\]
For $(1,0)$ to be stable fixed point, two conditions must be satisfied
\begin{eqnarray}
	det(J) = \lambda_{1}.\lambda_{2}...\lambda_{N} > 0\\
	Tr(J) = \lambda_{1} + \lambda_{2} + ... + \lambda_{N} < 0	
\end{eqnarray}
with $\lambda_{1},\lambda_{2},...,\lambda_{N}$ are the eigenvalues of $J$.
\par First we use Laplace expansion of the determinant of $J(det(J))$ with respect to the odd-numbered rows (i.e, the rows where only one element is 1 and all other elements are zero.) and the reduced determinant is a $N\times N$ determinant and it is expressed by
\[ det(J) = (-1)^{N}\begin{vmatrix}
a   &  \frac{K}{N} \epsilon   & \frac{K}{N} \epsilon & \cdots  & \cdots & \frac{K}{N} \epsilon\\\\
\frac{K}{N} \epsilon & a   & \frac{K}{N} \epsilon & \cdots  & \cdots & \frac{K}{N} \epsilon\\\\
\frac{K}{N} \epsilon   & \frac{K}{N} \epsilon & a & \cdots  & \cdots & \frac{K}{N} \epsilon\\\\
\vdots & \vdots & \ddots &        &  \vdots &  \vdots \\\\
\vdots & \vdots &        & \ddots &  \vdots &  \vdots \\\\
\frac{K}{N} \epsilon   & \frac{K}{N} \epsilon & \cdots  & \cdots & \frac{K}{N} \epsilon & a
\end{vmatrix}\]
In the first step we add the remaining rows to the first row and then pull out constant out of the determinant.\\\\
$det(J) =$\\ 
\[(-1)^{N}[a + (N-1)\frac{K}{N} \epsilon]\begin{vmatrix}
1   &  1   & 1 & \cdots  & \cdots & 1\\\\
\frac{K}{N} \epsilon & a   & \frac{K}{N} \epsilon & \cdots  & \cdots & \frac{K}{N} \epsilon\\\\
\frac{K}{N} \epsilon   & \frac{K}{N} \epsilon & a & \cdots  & \cdots & \frac{K}{N} \epsilon\\\\
\vdots & \vdots & \ddots &        &  \vdots &  \vdots \\\\
\vdots & \vdots &        & \ddots &  \vdots &  \vdots \\\\
\frac{K}{N} \epsilon   & \frac{K}{N} \epsilon & \cdots  & \cdots & \frac{K}{N} \epsilon & a
\end{vmatrix}\]
In the next step we perform the row operation $R_{i} \rightarrow R_{i} -(\frac{K}{N} \epsilon) R_{1} $, where $R_{i}$ denotes $i$-th row with $i = 2,3,...N$.\\\\
$det(J) =$\\
\[(-1)^{N}[a + (N-1)\frac{K}{N} \epsilon]\begin{vmatrix}
1   &  1   & 1 & \cdots  & 1\\\\
0 & (a -\frac{K}{N} \epsilon)   & 0 & \cdots  & 0\\\\
0 & 0 & (a -\frac{K}{N} \epsilon) & \cdots   & 0\\\\
\vdots & \vdots & \ddots &        &  \vdots \\\\
\vdots & \vdots &        & \ddots &  \vdots \\\\
0 & 0   & 0 & \cdots   & (a -\frac{K}{N} \epsilon)
\end{vmatrix}\]
so,
\begin{equation}
	det(J) = (-1)^{N}[a + (N-1)\frac{K}{N} \epsilon](a -\frac{K}{N} \epsilon)^{N-1}
\end{equation}
or
\begin{equation}
	(-1)^{N}[a + (N-1)\frac{K}{N} \epsilon](a -\frac{K}{N} \epsilon)^{N-1} > 0
\end{equation}

Now putting value of $a$, we get $a + (N-1)\frac{K}{N} \epsilon = -1$. This implies that
\begin{eqnarray}
	(a -\frac{K}{N} \epsilon)^{N-1} < 0\\
	K\epsilon > -1
\end{eqnarray}
Hence the solid black curve in the $K-\epsilon$ phase space is a rectangular hyperbola having the equation $K\epsilon = -1$. For $K\epsilon > -1$, $(1,0)$ is a stable fixed point and for $K\epsilon < -1$, $(1,0)$ is unstable.\\
Again from equation(14)
\begin{eqnarray}
	Tr(J) < 0\\
	bN < 0\\
	K > - \frac{N\alpha}{N-1}
\end{eqnarray}
Equations (18) and (21) must be satisfied for $(1,0)$ to be a stable fixed point in the network.
Similarly we check the stability of $(-1,0)$ and find that it is unstable in the whole $K-\epsilon$ phase space.

\end{document}